\documentclass[twocolumn]{jpsj3}

\usepackage{amsmath}
\usepackage{graphicx}

\def\nn{\nonumber}

\title{Electron Wave Function in Armchair Graphene Nanoribbons}

\author{
Ken-ichi \textsc{Sasaki}$^{1}$,
Katsunori \textsc{Wakabayashi}$^{1,2}$,
and
Toshiaki \textsc{Enoki}$^{3}$
}

\inst{
$^{1}$ International Center for Materials Nanoarchitectonics, \\
National Institute for Materials Science, Namiki, Tsukuba 305-0044,
Japan
 \\
$^{2}$ PRESTO, Japan Science and Technology Agency,
Kawaguchi 332-0012, Japan \\
$^{3}$ Department of Chemistry, Tokyo Institute of Technology,
Ookayama, Meguro-ku, Tokyo 152-8551, Japan
}

\recdate{\today}

\abst{
 By using analytical solution of a tight-binding model for armchair
 nanoribbons,
 it is confirmed that the solution represents the standing wave 
 formed by intervalley scattering and 
 that pseudospin is invariant under the scattering.
 The phase space of armchair nanoribbon 
 which includes a single Dirac singularity is specified.
 By examining the effects of boundary perturbations on the wave
 function, 
 we suggest that the existance of a strong boundary potential 
 is inconsistent with the observation in a recent 
 scanning tunneling microscopy. 
 Some of the possible electron-density superstructure patterns
 near a step armchair edge located on top of graphite are presented.
 It is demonstrated that 
 a selection rule for the G band in Raman spectroscopy
 can be most easily reproduced with the analytical solution.
}

\kword{graphene, armchair edge, nanoribbon, STM, Raman spectroscopy,
intervalley scattering, pseudospin, superstructure, HOPG}

\begin{document}

\maketitle

\section{Introduction}

Graphene nanoribbons are particularly useful 
in studying the properties 
which are hidden in carbon nanotubes 
by the periodic boundary condition.~\cite{saito92apl}
The boundary of a nanoribbon consists of 
two symmetrical edge structures,
armchair and zigzag edges.
A transmission electron microscope image illustrates that 
the edge of a sample 
prepared by micromechanical
cleavage,~\cite{novoselov05pnas} 
is a mixture of armchair and zigzag edges.~\cite{gupta09}
Some attempts have been made to prepare 
a regular zigzag or armchair edge over a large length more than
20 nm.~\cite{jia09} 

It is known that 
the behavior of the electrons near zigzag edge 
differs greatly from that near armchair edge in many respects. 
For example, 
armchair edge gives rise to intervalley scattering,
while zigzag edge produces intravalley scattering.
This originates from the fact that 
the graphene Brillouin zone in the $k$-space 
is given by rotating the hexagonal unit cell in real space 
by 90$^\circ$.~\cite{sasaki10-forward} 
On the other hand, 
zigzag edge changes the orientation of pseudospin, 
while armchair edge does not.
This behavior of pseudospin results from that 
only A or B (both A and B) sublattice appears at zigzag (armchair) edge.
These differences between zigzag and armchair edges 
cause observable effects to appear in Raman spectra. 
Because the D band in Raman spectra is relevant to intervalley
scattering, it follows that the D band intensity is enhanced  
by armchair edge.~\cite{canifmmode04sec}
The behavior of pseudospin is essential to 
a selection rule of the G band 
for the graphene edge.~\cite{sasaki10-jpsj}

Besides the differences 
in the behavior of the scattering and pseudospin,
armchair and zigzag edges are distinct 
with respect to the number of Dirac singularities.
In the case of nanotube, 
there are two Dirac singularities 
corresponding to the K and K$'$ points
at the corners of the first Brillouin zone. 
By means of an effective-mass theory, 
we discussed in a previous paper that
there is a single Dirac singularity
in the phase space of armchair nanoribbon,
while there is no Dirac singularity 
in the phase space of the standing wave of zigzag nanoribbon.~\cite{sasaki10-forward} 
The existence of a single Dirac singularity 
in armchair nanoribbon 
is of particular interest in transport property 
because the existence of the Dirac singularity 
can suppress backward scattering.
The appearance of the edge states~\cite{tanaka87,fujita96} 
near zigzag edge is a consequence of 
the absence of the Dirac singularity for zigzag nanoribbon.

In this paper, 
by constructing analytical solution 
of a tight-binding model for armchair nanoribbons,
we account for the features of armchair edge,
such as intervalley scattering, the behavior of
pseudospin, and the single Dirac singularity 
in the phase space.
Since the tight-binding model has an atomic resolution, 
we can apply the solution to some recent experimental results 
obtained by scanning tunneling microscopy.~\cite{yang10,sakai10}

Here, we would like to mention 
the previously published literature on
analytical solution for the wave 
function in armchair nanoribbon. 
Compernolle {\it et al.}~\cite{compernolle03} 
constructed the wave functions in nanoribbons
(achiral nanotubes with edges) using the transfer matrix method. 
They applied the result to obtain the transmission coefficient. 
Zheng {\it et al.}~\cite{zheng07} conducted analytical study of
electronic state in armchair nanoribbons. 
They showed that all armchair nanoribbons acquire nonzero energy gap
due to the variation of hopping integral near the edges. 
The above-mentioned authors employed a simple nearest-neighbor
tight-binding Hamiltonian that was shown to be a good approximation for
describing the electrons in nanotubes and graphene systems.
In this paper, we use the same Hamiltonian to 
examine armchair nanoribbons.

This paper is organized as follows.
In \S\ref{sec:wf} 
we solve a simple tight-binding model 
for armchair nanoribbons,
and obtain the energy dispersion eq.~(\ref{eq:disp})
and the wave function eq.~(\ref{eq:fwf}). 
The characteristic features of the wave function are pointed out.
By using eq.~(\ref{eq:fwf})
we will show in \S\ref{ssec:STM} that 
a feature of the Dirac singularity for armchair edge has been 
observed in recent scanning tunneling microscopy experiment by 
Yang {\it et al.}~\cite{yang10}, and in \S\ref{ssec:gband}
that the electron-phonon matrix element 
for the G band in Raman spectra 
which we have derived numerically with a tight-binding model 
in a previous paper~\cite{sasaki09} 
can be derived analytically.
Discussion and summary are given in \S\ref{sec:ds}.

\section{Construction of Wave Function}\label{sec:wf}

\begin{figure}[htbp]
 \begin{center}
  \includegraphics[scale=0.5]{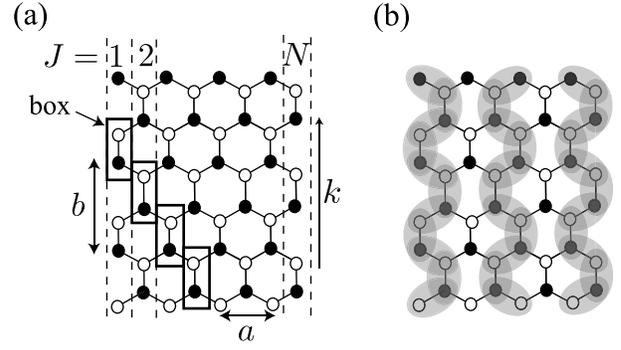}
 \end{center}
 \caption{
 (a) The structure of an armchair graphene nanoribbon. 
 The index $J$ takes from $1$ to $N$.
 The lattice constant is $a$ ($=\sqrt{3}a_{\rm cc}$).
 The unit length and the wave vector along the axis 
 are denoted by $b$ ($=3a_{\rm cc}$) and $k$, respectively. 
 Carbon atoms are divided into A ($\bullet$) and B ($\circ$) atoms.
 (b) The electron-density pattern of the Dirac singularity state in
 a metallic armchair nanoribbon with $N=8$.
 See also the STM topography shown in Fig.~3 of
 Ref.~\citen{yang10}.
 }
 \label{fig:armribbon}
\end{figure}

Let $\phi_{A}^{J}$ and $\phi_{B}^{J}$ 
be the probability amplitudes of the A and B-atoms
at the box in $J$th line shown in Fig.~\ref{fig:armribbon}(a),
then the Schr\"odinger equation for the electron in an armchair
nanoribbon is written as the recurrence equation,
\begin{align}
 -\epsilon 
 \begin{pmatrix}
  \phi_{A}^{J} \cr
  \phi_{B}^{J}
 \end{pmatrix}
 =&
 \begin{pmatrix}
  0 & 1 \cr
  1 & 0
 \end{pmatrix}
 \begin{pmatrix}
  \phi_{A}^{J} \cr
  \phi_{B}^{J}
 \end{pmatrix}
 +
 \begin{pmatrix}
  0 & e^{-ikb} \cr
  1 & 0 
 \end{pmatrix}
 \begin{pmatrix}
  \phi_{A}^{J-1} \cr
  \phi_{B}^{J-1}
 \end{pmatrix}
 \nn \\
 &+
 \begin{pmatrix}
  0 & 1 \cr
  e^{+ikb} & 0 
 \end{pmatrix}
 \begin{pmatrix}
  \phi_{A}^{J+1} \cr
  \phi_{B}^{J+1}
 \end{pmatrix}.
 \label{eq:ee}
\end{align}
Here, $\epsilon$ is the energy eigenvalue in units of the hopping
integral ($\gamma=3$ eV), $k$ is the wave vector along the edge, and $b$ 
($\equiv 2l$) denotes the unit length along the axis of armchair
nanoribbon [see Fig.~\ref{fig:armribbon}(a)].
In obtaining eq.~(\ref{eq:ee}) 
we have used the Bloch theorem, by which 
the amplitude of the B-atom in $J-1$th line
that is located nearest to the A-atom at the box 
in $J$th line 
is given by $e^{-ikb} \phi_{B}^{J-1}$.
Note that $J$ in eq.~(\ref{eq:ee}) takes from $J=2$ to $N-1$.
For $J=1$ and $J=N$, the recurrence equations are given by 
\begin{align}
 &
 -\epsilon 
 \begin{pmatrix}
  \phi_{A}^{1} \cr
  \phi_{B}^{1}
 \end{pmatrix}
 =
 \begin{pmatrix}
  0 & 1 \cr
  1 & 0
 \end{pmatrix}
 \begin{pmatrix}
  \phi_{A}^{1} \cr
  \phi_{B}^{1}
 \end{pmatrix}+
 \begin{pmatrix}
  0 & 1 \cr
  e^{+ik a} & 0 
 \end{pmatrix}
 \begin{pmatrix}
  \phi_{A}^{2} \cr
  \phi_{B}^{2}
 \end{pmatrix},
 \label{eq:bc-pre1}
 \\
 &
 -\epsilon 
 \begin{pmatrix}
  \phi_{A}^{N} \cr
  \phi_{B}^{N}
 \end{pmatrix}
 = 
 \begin{pmatrix}
  0 & 1 \cr
  1 & 0
 \end{pmatrix}
 \begin{pmatrix}
  \phi_{A}^{N} \cr
  \phi_{B}^{N}
 \end{pmatrix}+
 \begin{pmatrix}
  0 & e^{-ikb} \cr
  1 & 0 
 \end{pmatrix}
 \begin{pmatrix}
  \phi_{A}^{N-1} \cr
  \phi_{B}^{N-1}
 \end{pmatrix}.
\end{align}
When we assume that eq.~(\ref{eq:ee}) is satisfied for $J=1,\cdots,N$,
these equations at $J=1$ and $J=N$ can be included as the boundary
condition; 
\begin{align}
\begin{pmatrix}
 \phi^{0}_{A} \cr \phi^{0}_{B}
\end{pmatrix}
 = 0, 
 \ \
\begin{pmatrix}
 \phi^{N+1}_{A} \cr \phi^{N+1}_{B}
\end{pmatrix}
=0.
 \label{eq:bc}
\end{align}
In the following, we solve the recurrence equation of
eq.~(\ref{eq:ee}) with the boundary condition
of eq.~(\ref{eq:bc}).

First, let us introduce the following matrices,
\begin{align}
 K =
 \begin{pmatrix}
  \epsilon & 1 \cr
  1 & \epsilon 
 \end{pmatrix}, \ \ 
 T = 
 \begin{pmatrix}
  0 & 1 \cr z & 0
 \end{pmatrix}, \ \
 T^{-1} = 
 \begin{pmatrix}
  0 & z^* \cr 1 & 0
 \end{pmatrix},
 \label{eq:simp}
\end{align}
where $z=e^{+ikb}$, $z^*=e^{-ikb}$, and 
$kb \in [-\pi,\pi)$.
Then eq.~(\ref{eq:ee}) is rewritten as
\begin{align}
 T \phi^{J+1} +K \phi^{J} +T^{-1} \phi^{J-1} &= 0,\ \ (J=1,\cdots,N)
 \label{eq:eeM}
\end{align}
where we have defined $\phi^J \equiv {}^t(\phi^{J}_{A},\phi^{J}_{B})$.
By multiplying eq.~(\ref{eq:eeM}) by $T^{-1}$ from the left-hand side, 
we get $\phi^{J+1} +T^{-1}K \phi^{J} +(T^{-1})^2 \phi^{J-1} = 0$.
Because the matrix $T^{-1}$ satisfies 
$(T^{-1})^2 = z^* \sigma_0$,
where $\sigma_0$ is $2\times2$ identity matrix, 
the recurrence equation can be diagonalized by some matrix $U$ as 
\begin{align}
 \varphi^{J+1} + U^{-1} T^{-1}K U \varphi^{J} + \frac{1}{z} \varphi^{J-1}
 = 0,
 \label{eq:varphi}
\end{align}
where $\varphi^J \equiv U^{-1}¡¡\phi^J$.
It is straightforward to show that 
the eigenvalues of the matrix $T^{-1}K$
are given by 
\begin{align}
 \lambda_\pm = 
 \frac{1+z\pm \sqrt{(-1+z)^2+4z\epsilon^2}}{2z},
 \label{eq:ev}
\end{align}
and the corresponding eigen spinors are 
\begin{align}
 \phi_\pm = 
 \begin{pmatrix}
  \frac{1-z\pm \sqrt{(-1+z)^2+4z\epsilon^2}}{2z\epsilon} \cr 1
 \end{pmatrix}.
 \label{eq:eigenspinor}
\end{align}
The eigenvalue $\lambda_+$ (or $\lambda_-$) 
is a multivalued function 
since $\sqrt{z}$ has two branches: 
\begin{align}
 w_+ = + e^{ikl}, \ \
 w_- = - e^{ikl}.
\end{align}
The eigenvalue $\lambda_+$ with the $w_+$ ($w_-$) branch 
is equivalent to the eigenvalue $\lambda_-$ with the $w_-$ ($w_+$) branch.
As a result, we may consider the eigenvalue $\lambda_+$ only. 
Here we consider the $w_+$ branch.
We will show later that
the $w_-$ branch gives the same wave function as the $w_+$ branch, and
that these $w_+$ and $w_-$ branches correspond to different states
in the case of carbon nanotubes.

Next, we solve the characteristic equation 
with respect to eq.~(\ref{eq:varphi}) as 
\begin{align}
 & x^2 +\lambda_+ x + \frac{1}{z} = 0
 \nn \\
 & \leftrightarrow \left( x - w_+^{-1} e^{+i\theta} \right)
 \left( x - w_+^{-1} e^{-i\theta} \right) = 0,
\end{align}
where the parameter $\theta \in (0,\pi)$ 
is defined by $2\cos\theta  \equiv - w_+ \lambda_+$ or 
\begin{align}
 2\cos\theta 
 \equiv -
 \frac{w_+^{-1}+w_++\sqrt{z+z^*-2+4\epsilon^2}}{2}.
 \label{eq:cos}
\end{align}
The solution to the characteristic equation is given by 
$x = w_+^{-1} e^{+i\theta}$ or $x = w_+^{-1} e^{-i\theta}$.
Since $w_+^{-1}+w_+=2\cos(kl)$ and
$z+z^*=2\cos(2kl)$, eq.~(\ref{eq:cos}) leads to the 
energy dispersion relation,
\begin{align}
 \epsilon^2 = 1 + 4 \cos^2 \theta + 4 \cos\theta \cos(kl).
 \label{eq:disp}
\end{align}
By employing the standard method 
for solving the recurrence equation,
general $J$th term is written by 
the first and second terms, 
$\phi^1_+$ and $\phi^2_+$, as
\begin{align}
 \phi_+^J 
 = w_+^{-(J-2)}\frac{\sin(J-1)\theta}{\sin\theta} \phi^2_+
 - w_+^{-(J-1)} \frac{\sin(J-2)\theta}{\sin\theta} \phi^1_+.
 \label{eq:varphiJ}
\end{align}

To this point, we have not yet used the boundary condition.
The remaining procedure for obtaining the solution 
is applying the boundary condition to eq.~(\ref{eq:varphiJ}). 
The boundary condition of armchair edge
is given by $\phi^0=0$.
Thus, eq.~(\ref{eq:varphi}) with $J=1$ gives
$\phi^2=-T^{-1}K\phi^1$, so that
we have
\begin{align}
 \phi^2_+ 
 = -\lambda_+ \phi^1_+ 
 = \frac{2\cos\theta}{w_+} \phi^1_+.
 \label{eq:bc-ps}
\end{align}
By putting this into eq.~(\ref{eq:varphiJ}),
we obtain 
\begin{align}
 \phi^J_+
 = w_+^{-(J-1)}\frac{\sin J\theta}{\sin\theta}
 \phi^1_+.
 \label{eq:varphi+}
\end{align}
The eigen spinor $\phi_+^1$ may be represented 
in a simpler fashion as 
\begin{align}
 \phi^1_+ = 
 \begin{pmatrix}
  1+\frac{2\cos\theta}{w_+} \cr -\epsilon
 \end{pmatrix}=
 |\epsilon|
 \begin{pmatrix}
  e^{-i\Theta} \cr -s
 \end{pmatrix},
 \label{eq:ps}
\end{align}
where the variable $\Theta$ is a function of $k$ and $\theta$, 
and $s$ takes $+1$ ($-1$) for a conduction (valence) band state
[$\epsilon=s|\epsilon|$].
Therefore, the wave function is written as
\begin{align}
 \phi^J_+(s,k,\theta)
 = C e^{-ikl(J-1)}\frac{\sin J\theta}{\sin\theta}
 \begin{pmatrix}
  e^{-i\Theta(k,\theta)} \cr -s
 \end{pmatrix},
 \label{eq:bwf}
\end{align}
where $C$ is the normalization constant.

Several characteristic features of armchair nanoribbons
can be derived from the wave function eq.~(\ref{eq:bwf})
and energy dispersion relation eq.~(\ref{eq:disp}).
They are listed as follows.

(I) Armchair nanoribbons are metallic 
for the case that $N=3i-1$,~\cite{zhu98,sato99,wakabayashi99}
where $i$ is integer.
From the energy dispersion relation of eq.~(\ref{eq:disp}) 
one can see that the state with $\epsilon=0$
corresponds to the point $(0,2\pi/3)$ 
in the parameter space $(k,\theta)$.
The parameter $\theta$ is quantized 
as $\theta_n = n\pi/(N+1)$ [$n=1,\cdots,N$]
by imposing the boundary condition on the wave function:
$\phi^{N+1}_+(s,k,\theta)=0$.
The quantized parameter $\theta_n$ takes $2\pi/3$
for the case that $N=3i-1$ and $n=2i$.

(II) The angle $\Theta(k,\theta)$ in eq.~(\ref{eq:bwf})
is singular at the point $(0,2\pi/3)$.
By putting $(\delta k,2\pi/3+\delta \theta)$ into 
$1+2\cos\theta/w_+$ of eq.~(\ref{eq:ps}), 
we have $|\epsilon|e^{-i\Theta}=(-\sqrt{3}\delta\theta+i\delta k l)$
and $|\epsilon|=\sqrt{3\delta\theta^2+(\delta k l)^2}$.
The parameter $\Theta$ changes from $0$ to $2\pi$
when $(\delta k,\delta \theta)$ moves along a path 
enclosing the singularity point. 
This behavior of $\Theta$ results in a nontrivial Berry' phase
which is a necessary condition for an armchair nanoribbon
to exhibit a ballistic transport.~\cite{areshkin07,sasaki10-forward}

(III) The wave function 
can be understood as the standing wave 
formed by intervalley scattering.
This feature can be made more transparent 
by rewriting $\sin(J\theta)$ of eq.~(\ref{eq:fwf}) 
in terms of exponential functions as 
\begin{align}
 \phi^{J}_+(s,k,\theta) \propto 
 e^{-iklJ} \left[
 e^{+iJ\theta} 
 \begin{pmatrix}
  e^{-i\Theta} \cr -s
 \end{pmatrix} 
 - e^{-iJ\theta}
 \begin{pmatrix}
  e^{-i\Theta} \cr -s
 \end{pmatrix}
 \right],
\label{eq:scawf}
\end{align}
where the parameter $\theta \in (0,\pi)$ is a positive value.
The exponential function 
$e^{+iJ\theta}$ ($e^{-iJ\theta}$)
indicates the propagating wave in the direction of increasing (decreasing) $J$.
Hence, for the armchair edge at $J=1$,
the second term represents the incident wave to the edge and
the first term represents the reflected wave.
The momentum transfer through this scattering process,
$-\theta \to \theta$, is given by $2\theta$.
Since $\theta = 2\pi/3$ ($\equiv \theta_{\rm F}$)
is coincident with $k_{\rm F}a/2$, 
where $k_{\rm F}$ is the wave vector at the K point,
we can see that the change in the wave vector
at the armchair edge
is given by $2\theta_{\rm F}=k_{\rm F}a$.
This change of the wave vector corresponds to 
the intervalley scattering.

(IV) The spinors of the incident and edge reflected waves
in eq.~(\ref{eq:scawf}) are equal.
Hence, the pseudospin does not change 
under the intervalley scattering
at armchair edge.
This feature originates from the fact 
that both the A and B-atoms 
appear equivalently at the armchair edge.
It is also clear that 
the probability densities at the A and B-atoms 
in $J$th line are equal
and the electron-density pattern
depends only on the distance from the armchair edge.

(V) There is a single Dirac singularity in the system.
To make this point clear, 
we rewrite the standing wave eq.~(\ref{eq:scawf}) as
\begin{align}
 \phi^{J}_+ = \psi_{\rm K} - \psi_{\rm K'},
\end{align}
where $\psi_{\rm K}$ ($\psi_{\rm K'}$)
corresponds to a Bloch state near the K (K$'$) point.
The minus sign in front of $\psi_{\rm K'}$
shows that this standing wave 
is an anti-symmetric combination of the two Bloch states.
If the two states were independent of each other,
we may expect that the symmetric combination 
$\psi_{\rm K} +\psi_{\rm K'}$ is also a solution.
However, such symmetric solution should be excluded
because $\psi_{\rm K} +\psi_{\rm K'}$
is proportional to $\cos(J\theta)$ and 
does not satisfy the armchair boundary condition.
Since the armchair boundary condition selects only the 
antisymmetric solution,
the two Dirac points (K and K$'$) are 
no longer independent of each other in armchair nanoribbon.
Therefore, the number of Dirac singularities is reduced to 
one in this system. 
On the other hand, 
the two Dirac points (K and K$'$) are independent of each other
in the case of nanotube
because the symmetric and antisymmetric combinations 
can satisfy the periodic boundary condition of nanotube.

\begin{figure}[htbp]
 \begin{center}
  \includegraphics[scale=0.45]{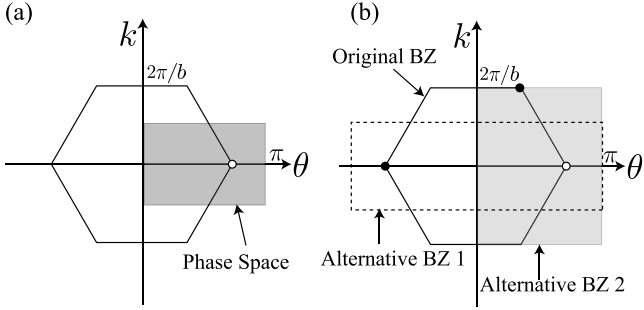}
 \end{center}
 \caption{
 (a) The region denoted by shadow, $\theta\in(0,\pi)$ and
 $k\in[-\pi/b,\pi/b)$, represents the phase space of armchair
 nanoribbon.
 The empty circle indicates the position of the Dirac singularity.
 (b) Two alternative Brillouin zones (BZ 1 and BZ 2) for graphene.
 The BZ 1 is useful to understand 
 that the phase space of armchair nanoribbon
 is given by folding BZ 1 into the positive region of $\theta$ with
 respect to the $k$-axis.
 The shadow region (BZ 2) is used to discuss on the Riemann sheets
 (see text).
 Note that alternative BZs contain two singularities
 corresponding to the K and K$'$ points, while 
 the phase space of armchair nanoribbon has a single singularity.
 }
 \label{fig:pspace}
\end{figure}

(VI) The phase space of armchair nanoribbon is given by 
$\theta\in(0,\pi)$ and $k\in[-\pi/b,\pi/b)$
[See Fig.~\ref{fig:pspace}(a)].
By comparing the phase space of armchair nanoribbon
with the usual hexagonal Brillouin zone (BZ) of graphene, 
it can be seen that 
the former covers only the half region of the latter.
To explore this feature further, 
we consider the other branch $w_-$.
Repeating analysis similar to that above, we obtain 
the corresponding energy dispersion relation and 
the wave function as
\begin{align}
 & 
 \epsilon^2 = 1 + 4 \cos^2 \theta' - 4 \cos\theta' \cos(kl),
 \label{eq:disn}
 \\
 & \phi^J_-
 = w_-^{-(J-1)}\frac{\sin(J\theta')}{\sin\theta'}
 \begin{pmatrix}
  1+\frac{2\cos\theta'}{w_-} \cr -\epsilon
 \end{pmatrix},
 \label{eq:varphi-}
\end{align}
where $\theta'\in (-\pi,0)$ is defined by
\begin{align}
 2\cos\theta' 
 \equiv -
 \frac{w_-^{-1}+w_- + \sqrt{z+z^*-2+4\epsilon^2}}{2}.
 \label{eq:cos'}
\end{align}
By subtracting eq.~(\ref{eq:cos'}) from eq.~(\ref{eq:cos}), 
we get $\cos\theta' = \cos\theta + \cos(kl)$.
Using this relationship between $\theta$ and $\theta'$, 
we see that the point $(k,\theta)=(0,2\pi/3)$ 
corresponds to $\theta'=-\pi/3$, 
for which $\phi^J_-$ reproduces $\phi^J_+$. 
Generally, by setting $\theta'= \theta-\pi$ in eq.~(\ref{eq:varphi-}),
we obtain the wave function $\phi^J_+$
of eq.~(\ref{eq:varphi+}).
Thus, 
the $w_-$ branch does not yield a state which is  
independent of the $w_+$ branch.
Hence, we may consider the $w_+$ branch only.
This observation allows us to interpret 
the phase space of armchair nanoribbon 
shown in Fig.~\ref{fig:pspace}(a)
as the region
given by holding the alternative BZ 1 of 
nanotube shown in Fig.~\ref{fig:pspace}(b) 
into the positive region of $\theta$.
In the remainder of the paper, 
we abbreviate $\phi^J_\pm(s,k,\theta)$ as
$\phi^J_s(k,\theta)$,
\begin{align}
 \phi^J_s(k,\theta)
 = C e^{-ikl(J-1)}\frac{\sin J\theta}{\sin\theta}
 \begin{pmatrix}
  e^{-i\Theta(k,\theta)} \cr -s
 \end{pmatrix}.
 \label{eq:fwf}
\end{align}

It is interesting to note that the $\omega_+$ and $\omega_-$ branches
correspond to different states in the case of nanotube.
To show this, 
we consider two Riemann sheets for the variable $z=e^{ikb}$.
Thus, we consider the region, $k \in (-2\pi/b,2\pi/b]$,
or the alternative BZ 2 in Fig.~\ref{fig:pspace}(b), 
and choose the branch $w=\sqrt{z}$ as a single valued function 
for the two Riemann sheets. 
From the energy dispersion relation eq.~(\ref{eq:disp}) 
we see that 
two points in the Riemann sheets,
$(0,2\pi/3)$ and $(2\pi/b,\pi/3)$, result in $\epsilon=0$.
Note that 
$k=0$ and $k=2\pi/b$ belong to the different Riemann sheets. 
Hence, these points correspond to the independent 
Dirac singularities at the corners of the first BZ of a nanotube.

\section{Applications}\label{sec:app}

In \S\ref{ssec:STM},
we examine recent experimental result
of scanning tunneling microscopy (STM)
by using the analytical solution eq.~(\ref{eq:fwf}).

\subsection{STM Topography and Dirac Singularity}\label{ssec:STM}

The spatial dependence of the wave function $\phi^J_s(k,\theta)$
is given by $e^{-iklJ}\sin (J\theta)$.
Hence, the probability density 
$|\phi^J_s(k,\theta)|^2$
is proportional to $\sin^2 (J\theta)$ 
and exhibits a spatial oscillation.
Note that the probability density
does not depend on the wave vector $k$.
As a result, a STM image should be stable
against a change in the wave vector $k$ caused by, 
for example, a change of the bias voltage.

As we have seen, the Dirac singularity 
corresponds to the state with $\theta = 2\pi/3$ and $k=0$.
Therefore, the probability density for a state 
near the Dirac singularity with $\theta = 2\pi/3$
disappears at $J=3i$, where $i$ is integer,
as shown in Fig.~\ref{fig:armribbon}(b).
This behavior of the periodic absence of probability density
(line nodes) has been recognized in
literature.~\cite{tanaka87,treboux99,sato99,zheng07}
On the other hand, 
for a state satisfying $\theta = 2\pi/3+\delta \theta$,
that is, for a state away from the Dirac singularity,
the densities at $J=3i$ behave according to 
$\sin^2 (3i\delta\theta)$.
Hence, as $i$ increases or as the distance from armchair edge increases,  
the density at $J=3i$ increases gradually.~\cite{zheng07}
Consequently, 
the behavior of the periodic absence of the probability density
at $J=3i$ is not seen for the case that $i$ is sufficiently large 
or away from the armchair edge.

In the STM topography obtained by Yang {\it et al.}
[Fig.~3 of Ref.~\citen{yang10}],
the density at $J=3i$ is suppressed for the case $i \le 3$
and the appreciable density appears for the case $i>3$.
This feature of the STM topography 
allows us to interpret 
the character of their sample
in the following way:
the state observed by their STM was a state away from the Dirac
singularity, that is, a state with $\delta \theta\ne 0$.
If we assume that the probability density at 
$J=3i$ with $i=4$ is comparable to that at $J=1$, 
we have the following equality determining $\delta\theta$,
\begin{align}
 \sin^2(12\delta \theta) = \sin^2(2\pi/3).
\end{align}
The minimum value satisfying this equation 
is $\delta \theta = \pi/36$.
This value of $\delta\theta$ 
corresponds to the states with energy 
$|E| \approx \gamma \sqrt{3}|\delta \theta| = 0.45$ eV.
Note that the energy value coincides with the energy difference between
the Dirac point and the Fermi energy for graphene on 6H-SiC(0001)
as pointed out in Ref.~\citen{yang10}.
Note also that a small $\delta \theta$ 
gives rise to a long-range electron-density oscillation 
of length $(a/2)(\pi/\delta\theta)/3$.
A long-range oscillation with a period of about 2.5 nm 
is actually observed by Yang {\it et al.}
The value of $\delta \theta$ which reproduces 2.5 nm 
is about $\pi/60$. 
The actual period of a long-range oscillation 
may depend on the details of the substrate and 
the perturbations at the boundary.

Here, we examine the effects of boundary perturbations, such as
a change in bond length or a change in on-site potential, 
on the electronic state.
First, 
suppose that the hopping integral between the A and B-atoms 
at $J=1$ differs from that at $J\ne 1$ by $\delta \gamma$.
Then the first order energy shift is given from eq.~(\ref{eq:fwf})
by
\begin{align}
 \Delta \epsilon 
 \equiv 
 [\phi_s^1]^\dagger \left(\delta \gamma\sigma_x \right) \phi_s^1
 = -2 s \delta \gamma  |C|^2 \cos \Theta.
\end{align}
Because the energy shift is proportional to $s$,
a highest energy state in the valence band
and a lowest energy state in the conduction band
undergo the opposite energy shift. 
This means that the energy gap changes. 
For example, an armchair nanoribbon with $N=3i-1$
acquires a finite energy gap
due to a change in bond length at the boundary.~\cite{zheng07}
Since $\Delta \epsilon \propto \cos \Theta$,
the gap is actually produced as a consequence of 
the shift, $\delta \theta$, of the Dirac
singularity point along the axis of $\theta$.
Fujita {\it et al.}~\cite{fujita97} 
showed by a numerical calculation that 
a nonzero $\delta \gamma$, that is, a nonzero $\delta \theta$ 
can be induced by Peierls instability 
for the case that $N$ is very small like 3 or 6.
Since $N$ is large in the experiment,
the contribution of the Peierls instability to $\delta\theta$
may be ignored.

Next, we consider that the on-site potential at $J=1$ 
differs from that at $J\ne 1$ by $v$.
The first order energy shift 
due to the boundary potential
is given by
\begin{align}
 \Delta \epsilon 
 \equiv 
 [\phi_s^1]^\dagger \left( v\sigma_0 \right) \phi_s^1
 = 2 v |C|^2.
\end{align}
Because the energy shift is independent of $s$,
a highest energy state in the valence band
and a lowest energy state in the conduction band
undergo the same energy shift. 
This means that a boundary potential does not change the energy gap.
However, this fact does not necessarily mean that 
the boundary potential does not produce an observable effect
in STM topography. 
The wave function can be sensitive to
the presence of the boundary potential.
In fact, in the presence of $v$, 
the recurrence equation eq.~(\ref{eq:bc-pre1})
is modified as $\phi^2 = -T^{-1}(K+v \sigma_0)\phi^1$,
so that we obtain
\begin{align}
 \phi^2_+ = -\left( \lambda_+ 
 + v T^{-1} \right)\phi^1_+,
\end{align}
instead of eq.~(\ref{eq:bc-ps}).
By putting this new boundary condition into eq.~(\ref{eq:varphiJ}),
we obtain the modified wave function,
\begin{align}
 \phi^J_+
 &= w_+^{-(J-1)}\frac{\sin J\theta}{\sin\theta} \phi^1_+ \nn \\
 &- w_+^{-(J-2)}\frac{\sin (J-1)\theta}{\sin\theta}
 vT^{-1} \phi^1_+.
 \label{eq:modwf}
\end{align}
When $v$ is sufficiently large in this equation, 
only the last term is important.
By using the definition of $T^{-1}$ in eq.~(\ref{eq:simp})
the last term can be rewritten as
\begin{align}
 - w_+^{-(J-1)} \frac{\sin (J-1)\theta}{\sin\theta}
 v
 \begin{pmatrix}
  0 & w_+^{-1} \cr
  w_+ & 0
 \end{pmatrix}
 \phi^1_+.
\end{align}
It vanishes at $J=1$ for any $\theta$, which 
shows a change of the boundary condition.
Note that the last term vanishes at $J=3i+1$ 
for a state near the Dirac singularity ($\theta=2\pi/3$),
while the original wave function vanishes at $J=3i$
for the state.
It is also interesting to note that 
the boundary potential can change the direction of pseudospin.

It is difficult to predict the value of boundary potential $v$, 
because there are a number of factors that can change the value of $v$.
For example, $v$ depends on functional group attached to 
the carbon atoms at the boundary.
However, some information about $v$ 
can be derived from an experimental STM image.
To this end, we consider the amplitude at $J=2$.
Then, for a state near the Dirac point ($\theta=2\pi/3$),
we obtain from eq.~(\ref{eq:modwf}) that 
\begin{align}
 \phi^2_+ \propto \phi^1_+ + vT^{-1} \phi^1_+.
\end{align}
If $|v|={\cal O}(1)$, 
a strong interference between the first and second terms 
can be expected in general. 
There is a possibility that 
the interference effect results in a strong difference 
between the electron-densities at $J=1$ and $J=2$.
Since the STM image of Yang {\it et al.}
[Fig.~3 of Ref.~\citen{yang10}]
shows clearly that the electron density at $J=2$
is similar to that at $J=1$, which indicates that
$|v|$ is much smaller than ${\cal O}(1)$.
On the other hand, 
the existence of a small boundary potential is reasonable
because a boundary potential about $|v|\approx 0.1$
can be induced by intrinsic perturbations, such as 
next nearest-neighbor hopping integral~\cite{sasaki09-hd} and phonon.
Detecting in STM images the signal of boundary potential $v$ 
is an interesting subject.

\subsection{Raman G Band}\label{ssec:gband}

\begin{figure}[htbp]
 \begin{center}
  \includegraphics[scale=0.45]{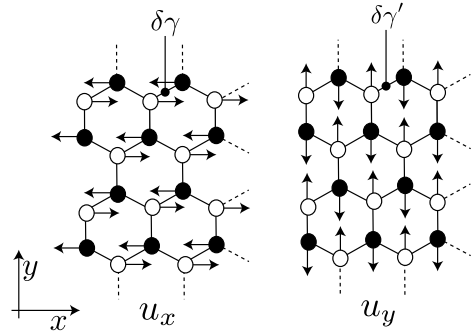}
 \end{center}
 \caption{
 The displacement vectors, $u_x$ and $u_y$, of the $\Gamma$ point
 optical phonon modes. 
 $\delta \gamma$ ($\delta \gamma'$) represents a change of the
 nearest-neighbor hopping integral caused by the vibration.
 }
 \label{fig:gband}
\end{figure}

The analytical wave function is useful 
in calculating the electron-phonon (el-ph) matrix element
for a process observed in Raman spectroscopy. 
Let us consider the G band in Raman spectra.
It is known that two optical phonon modes
at the $\Gamma$ point, whose displacement vectors 
are denoted by $u_x$ and $u_y$ shown in Fig.~\ref{fig:gband},
contribute to the G band.
The displacement vector $u_y$ is parallel to the armchair edge 
and $u_x$ is perpendicular to it.
Here, we calculate the el-ph matrix elements
for the $u_x$ and $u_y$ modes.

We define $G_x$ and $G_y$ as follows,
\begin{align}
\begin{split}
 \frac{G_x}{\delta \gamma} 
 &= \sum_{J=1}^N (\phi^{J})^\dagger \left[
 T\sigma_z
 \phi^{J+1} 
 + 
 \sigma_z T^{-1}
 \phi^{J-1} \right] +{\rm c.c.},
 \\
 \frac{G_y}{\delta \gamma'} 
 &= \sum_{J=1}^N (\phi^{J})^\dagger 
 \left[
 \frac{1}{2} \left\{
 T
 \phi^{J+1} 
 + 
 T^{-1}
 \phi^{J-1} 
 \right\}
 -\sigma_x
 \phi^{J} 
 \right] +{\rm c.c.}
\end{split}
\label{eq:Mxy}
\end{align}
It is straightforward to show that 
$G_x$ and $G_y$ are the expectation values of 
the perturbations caused by the lattice displacements $u_x$ and $u_y$,
respectively. 
Here, $\delta \gamma$ and $\delta \gamma'$
are the changes of the hopping integral
for the C-C bonds denoted in Fig.~\ref{fig:gband},
produced by the lattice displacements $u_x$ and $u_y$, respectively.
By putting $\phi^J_s$ of eq.~(\ref{eq:fwf}) 
into $\phi^J$ of eq.~(\ref{eq:Mxy}),
we obtain 
\begin{align}
\begin{split}
 & G_x=0, \\
 & \frac{G_y}{\delta \gamma'} =
- 3 \langle \sigma_x \rangle - \epsilon,
\end{split}
\label{eq:matG}
\end{align}
where
$\langle \sigma_x \rangle$ 
is the expectation value of the Pauli matrix $\sigma_x$
defined by 
\begin{align}
 \langle \sigma_x \rangle \equiv 
 \sum_{J=1}^N (\phi^{J}_s)^\dagger \sigma_x \phi^{J}_s
 =-s \cos\Theta.
\end{align}
In eq.~(\ref{eq:matG}),
we have obtained $G_y$ by putting 
the energy eigen equation of 
eq.~(\ref{eq:eeM}) with $K=\epsilon \sigma_0 + \sigma_x$
into the right-hand side of $G_y$ in eq.~(\ref{eq:Mxy}).

We note that this result of $G_y$ derived by using 
tight-binding model is slightly different from the result 
of an effective-mass theory.
The effective-mass theory gives~\cite{sasaki10-jpsj}
\begin{align}
 \frac{G_y}{\delta \gamma'} =- 3 \langle \sigma_x \rangle.
\end{align}
The difference between the two schemes is $\epsilon$.
Since $\epsilon$ is scaled by the hopping integral,
$\epsilon$ vanishes in the continuum limit, so that
this $\epsilon$ term does not appear in the effective mass theory. 
Here, we consider two direct consequences of the $\epsilon$ term.
First, because the Raman intensity is proportional to $G_y^2$,
the $\epsilon$ term may give rise to a small dependence 
of the Raman intensity on the laser energy.
The dependence about 10\%$/$eV is acceptable
for the process that satisfies $|\langle \sigma_x \rangle| \approx 1$.
Second, the strength of $G_y$ decreases 
for the state near the top (bottom) at the conduction (valence)
band. 
Thus, the Raman intensity for, if any, such a high energy state 
should be suppressed strongly.

The result in eq.~(\ref{eq:matG}) means that 
the $u_x$ mode is not Raman active mode near the armchair edge
and only the $u_y$ mode can be Raman active.
This is a phenomenon peculiar to armchair
edge~\cite{sasaki09,sasaki10-jpsj}
because it can be shown that 
the $G_x$ and $G_y$ do not vanish
in the case of a periodic graphene as
\begin{align}
\begin{split}
 & \frac{G_x}{\delta \gamma}=s\left(\cos(\theta+\Theta) - \cos(\theta-\Theta+kb)\right), \\
 & \frac{G_y}{\delta \gamma'} =
- 3 \langle \sigma_x \rangle - \epsilon.
\end{split}
\end{align}
Recent experiments for the G band Raman spectroscopy for graphene edge
by Cong {\it et al.}~\cite{cong10} and Begliarbekov {\it et al.}~\cite{begliarbekov10}
support this prediction.

\section{Discussion and Summary}\label{sec:ds}

\begin{figure}[htbp]
 \begin{center}
  \includegraphics[scale=0.4]{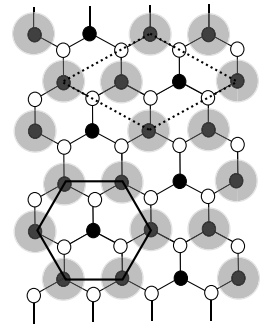}
 \end{center}
 \caption{
 The electron-density pattern of a low-energy state near the armchair
 step edge located on top of a HOPG.
 The unit of $(\sqrt{3}a\times\sqrt{3}a)\ 30^\circ$ superstructure
 is shown by the rhombus.
 This unit cell is the same as that of the Kekul\'e pattern.
 }
 \label{fig:super}
\end{figure}

It is often said that 
a step edge located on top of a highly oriented pyrolytic graphite (HOPG)
can have a regular structure over a large length more than
10 nm.~\cite{giunta01} 
The wave function for such a step edge 
can also be constructed analytically.
The equation for a step armchair edge 
on top of HOPG is given by replacing the matrix $K$ in eq.~(\ref{eq:eeM})
with $K-m\sigma_z$.
Here, $m$ represents the potential difference 
between the A and B sublattice which is induced by 
the stack in the Bernal configuration.
It is easy to show that
the spinor eigenstate of eq.~(\ref{eq:eigenspinor})
is modified by the presence of $m$ as 
\begin{align}
 \phi_+ = 
 \begin{pmatrix}
  \frac{1-z+ \sqrt{(-1+z)^2+4z(\epsilon^2-m^2)}}{2z(\epsilon-m)} \cr 1
 \end{pmatrix}.
\label{eq:mass}
\end{align}
Note that the lowest energy state with $k=0$ ($z=1$)
in the conduction band 
has amplitude only on the A-atoms for the case that $m>0$.
It is obvious that by taking the limit $\epsilon \to m$
in eq.~(\ref{eq:mass}), the normalized spinor is written as
\begin{align}
 \phi_+ =
 \begin{pmatrix}
  1 \cr 0
 \end{pmatrix},
\end{align}
which is a pseudospin polarized state.
The electron-density pattern of this state is shown in
Fig.~\ref{fig:super}, where
the density forms the hexagonal pattern.
It is important to recognize that 
the appearance of the hexagonal pattern results from two causes: 
(1) the asymmetry of the potential energies at the A and B sublattice
and (2) the existence of armchair edge.
The condition (1) is satisfied in the bulk of
HOPG~\cite{tapaszto08}
and epitaxial graphene on SiC(0001)~\cite{rutter07}.
The STM images show a triangular lattice there because
the amplitude can be observed only at one of the two sublattice
for the case that $m\ne 0$.
In addition to (1), near the armchair edge (2), 
the low energy electron does not have amplitude at the sites $J=3i$ 
[see Fig.~\ref{fig:armribbon}(b)] due to $\sin(J\theta)$.
As a result, a low bias STM topography observes 
the hexagonal electron-density pattern
near a step armchair edge on top of a HOPG.~\cite{niimi06,sakai10}

Beside the hexagonal pattern,
an electron-density pattern which is 
referred to as $(\sqrt{3}a\times\sqrt{3}a)\ 30^\circ$ 
or $(\sqrt{3}\times\sqrt{3})R\ 30^\circ$ superstructure
is frequently observed near defects on graphite, 
such as metal particles,~\cite{shedd92,valenzuela-benavides95}
isolated adsorbed molecules,~\cite{mizes89,kelly00} 
step edges,~\cite{albrecht88,giunta01,tapaszto08}
and lattice vacancies.~\cite{coratger92,yan94,ruffieux05,tapaszto08}
This superstructure 
is denoted by the rhombus in Fig.~\ref{fig:super}
(see ref.~\citen{shedd92}). 
The periodicity of the $(\sqrt{3}a\times\sqrt{3}a)\ 30^\circ$
superstructure is actually included in the wave function
for armchair edge as shown in Fig.~\ref{fig:super}.
Note, however, that the origin of this superstructure 
observed in the experiments for step edges
can not be attributed only to armchair edge since
the presence of zigzag edge might play an important role.
A simulation by Niimi {\it et al.}~\cite{niimi06} 
indicates such a possibility.
To fully specify the origin of 
the $(\sqrt{3}a\times\sqrt{3}a)\ 30^\circ$ superstructure,
it is desirable to construct analytical wave function 
for a mixed edge consisting of zigzag and armchair edge parts.

In summary,
we have constructed the analytical solution of the tight-binding model 
for armchair nanoribbon.
The wave function is described by the superposition of the incident
wave to the armchair edge and the scattered wave, i.e., 
the formation of standing wave.
The scattering process is intervalley scattering, in which the
pseudospin is invariant. 
The difference between armchair nanoribbon and
nanotube appears 
with respect to the number of Dirac singularities in their phase
spaces.
Since the armchair boundary condition identifies the half of graphene BZ
(including the K point)
and the other half of the graphene BZ
(including the K$'$ point), 
the phase space of armchair nanoribbon is given by the half region of
the graphene BZ.
As a result, in the phase space of armchair nanoribbon, there is a
single Dirac singularity whose wave function is written by the 
antisymmetric combination of the two states at the K and K$'$ points. 
We have examined the recent STM topography
by using the analytic solution.
It seems that the STM image obtained by 
Yang {\it et al.}~\cite{yang10} corresponds 
to a state away from the Dirac singularity.
This speculation is consistent with the appearance of a long-range
oscillation in their STM image.
Moreover, the observed STM image does not allow assuming
the existence of a strong boundary potential of order of the hopping
integral.
A lattice distortion at armchair edge 
can be taken into account as a shift of the Dirac singularity point
along the $\theta$-axis, 
which is similar to the curvature effect 
in metallic carbon nanotubes. 
In addition,
the analytical solution allows us to observe easily that 
$u_x$ mode is not Raman active near the armchair edge.
The analytical solution simplifies greatly 
the analysis which was done 
numerically with the tight-binding model,~\cite{sasaki09}
and is also useful for confirming the result obtained analytically 
with the effective-mass model~\cite{sasaki10-jpsj}.

\section*{Acknowledgments}

This work is supported by 
a Grant-in-Aid for Specially Promoted Research
(No.~20001006) from the Ministry of Education, Culture, Sports, Science
and Technology.

\bibliographystyle{./jpsj}

\end{document}